\documentclass{DISproc}
\usepackage{multirow}

\begin{document}
\title{Constraining the Intrinsic Heavy Quark PDF via Direct Photon Production in Association with a Heavy Quark Jet.}

\author{{\slshape K.Kova\v{r}\'{\i}k$^1$, T.Stavreva$^2$}\\[1ex]
$^1$Institute for Theoretical Physics, Karlsruhe Institute of Technology, Karlsruhe, 76128,Germany\\
$^2$Laboratoire de Physique Subatomique et de cosmologie, Grenoble, 38026, France}

\contribID{250}

\doi  

\maketitle

\begin{abstract}
	We investigate a possible use of direct photon production in
	association with a heavy quark to test different models of 
	intrinsic heavy quark parton distribution function (PDF)
	at the Tevatron, at the large hadron collider (LHC) and at RHIC.
\end{abstract}

\section{Introduction}
\begin{figure}[htb]
  \centering
  \includegraphics[width=0.49\textwidth]{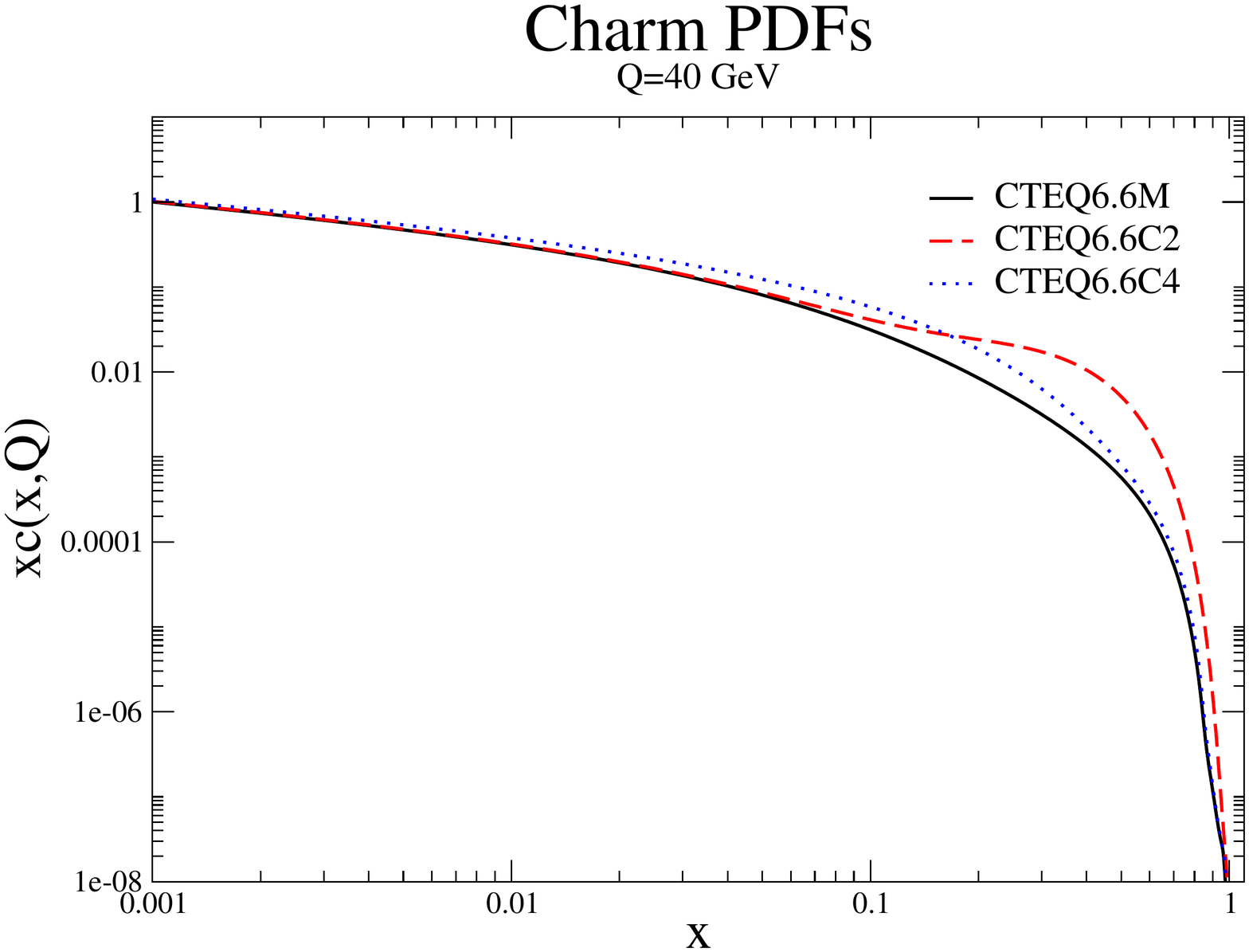}
  \includegraphics[width=0.49\textwidth]{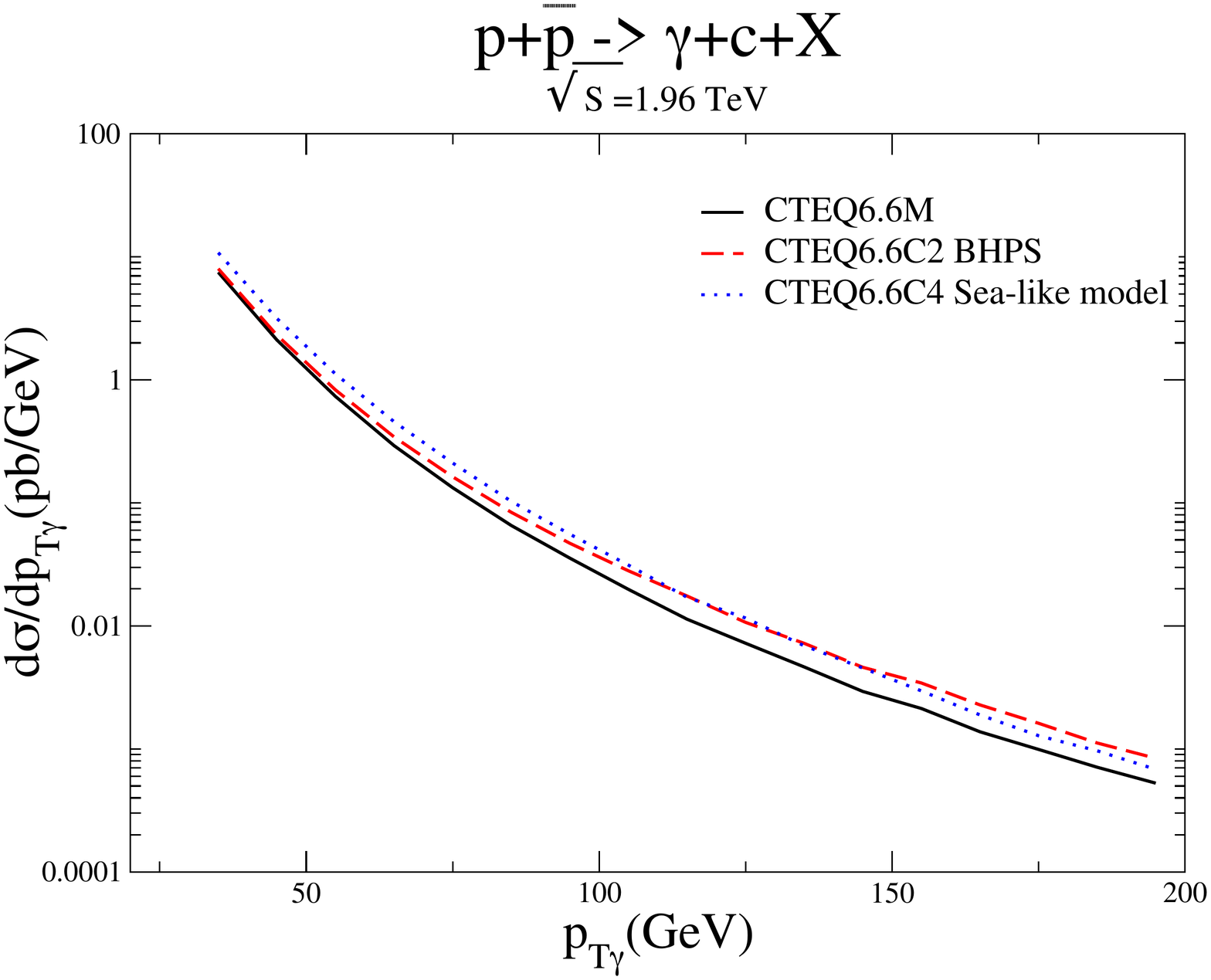}
  \caption{Charm parton distribution function $c(x,Q)$ at $Q=40$ GeV and the differential cross-section for the direct 
photon production with a charm quark jet.}\label{fig:charmPDF}
\end{figure}

Parton distribution functions (PDFs) are an essential component of any prediction involving colliding hadrons. 
In view of their importance, the proton PDFs have been a focus of long and dedicated global analyses performed 
by various groups \cite{Ball:2009mk, Martin:2009iq, Nadolsky:2008zw}. Most PDF analyses assume five active quark 
flavours with corresponding parton distributions $u(x,Q^2),d(x,Q^2),c(x,Q^2),s(x,Q^2)$ and $b(x,Q^2)$ but only
the light quark parton distribution functions $u,d$ and $s$ are non-zero at the input scale $Q_0$. The massive 
quark PDF are normally generated only radiatively from the gluon through a radiation of a quark-anti-quark pair.
This however does not have to be the case and there exist models predicting an intrinsic contribution to the charm
quark PDF \cite{Pumplin:2007wg}. Depending on the model, the intrinsic component changes the charm PDF at a higher 
scale as shown in Fig.~\ref{fig:charmPDF}. The BHPS model (shown as CTEQ6.6C2 in Fig.~\ref{fig:charmPDF}) is a light-cone 
model which enhances the charm PDF at large values of $x$ as opposed to the sea-like model (shown as CTEQ6.6C4 in 
Fig.~\ref{fig:charmPDF}) where the charm intrinsic contribution is proportional to the light flavour sea quark 
distributions and modifies the radiative charm PDF at all values of $x$.

The direct photon production in association with a heavy quark jet is one of only a few processes one can use to probe
a possible intrinsic heavy quark component at high-$x$. As is shown in Fig.~\ref{fig:charmPDF} for the direct photon
production process at the Tevatron \cite{Stavreva:2009vi}, the different models of intrinsic charm modify the 
prediction for this specific process at high transverse momenta of the photon. We discuss similar predictions 
for different collider scenarios.
\section{Direct photon production}
Single direct photons have long been considered an excellent probe of the structure of the proton due to 
their point-like electromagnetic coupling to quarks and due to the fact that they escape confinement.  
Concentrating on a double inclusive production of a direct photon with a heavy quark allows access to different PDF components.
Single direct photons couple mostly to valence quarks in the proton. By investigating direct photons accompanied by
heavy quark jets, one gains access to the gluon and the heavy quark PDF. That is because, at leading order ${\cal O}(\alpha\alpha_s)$, 
the direct photon with a heavy quark arises only from $g Q \to \gamma Q$ Compton scattering process as opposed to 
the single photon in which case a Compton scattering contribution $gq \to \gamma q$ competes against a contribution from quark 
annihilation $q \bar q \rightarrow \gamma g$.

At leading order, we see that the initial state for the direct photon production with a heavy quark jet depends only on 
gluon PDF and heavy quark PDF where also the latter is often radiatively generated from the gluon leading to even stronger 
dependence of this process on the gluon PDF.

At next-to-leading order (NLO), the dependence of the direct photon production even with a heavy quark jet lessens due to
\begin{figure}[htb]
  \centering
  \includegraphics[width=0.45\textwidth]{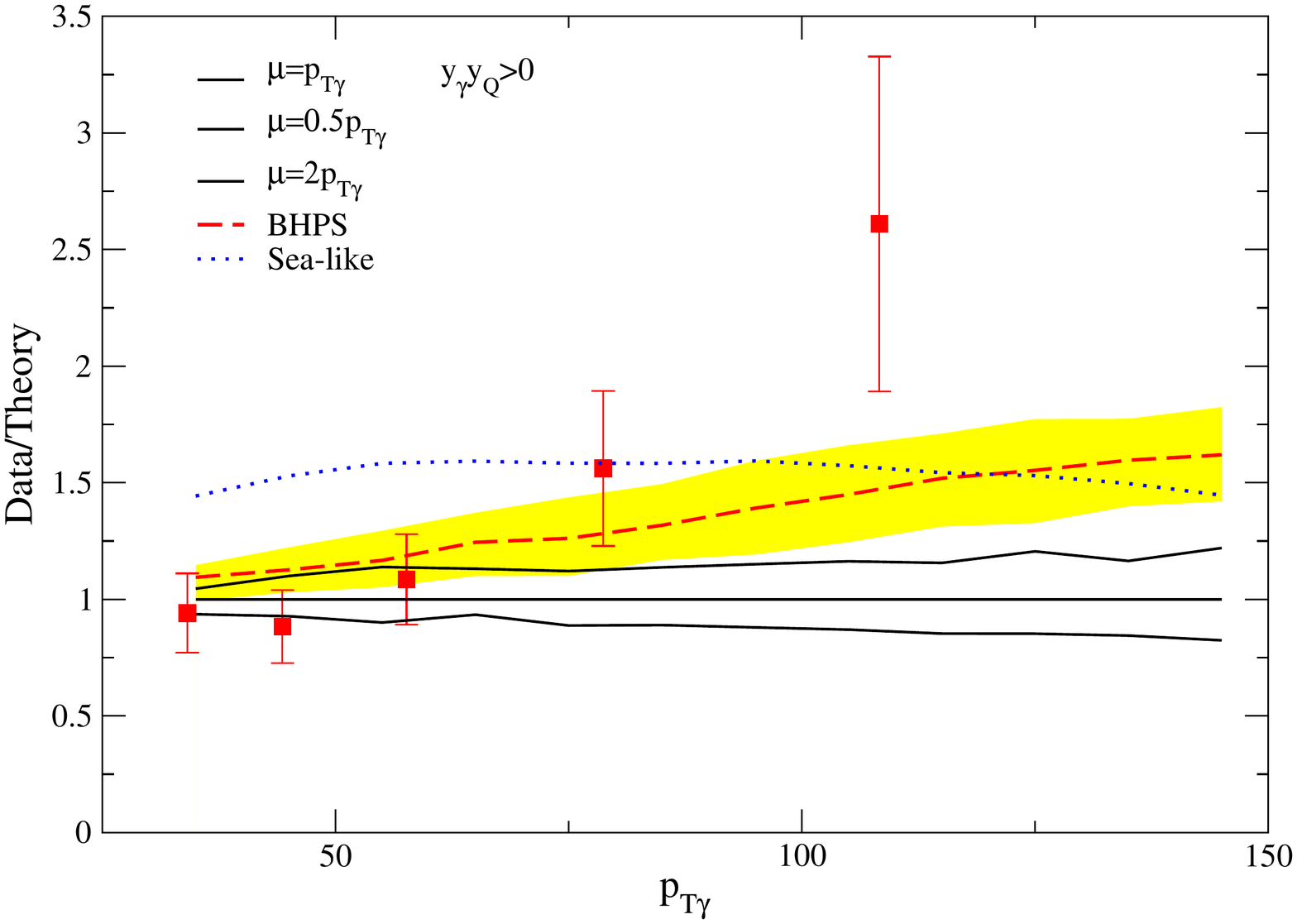}
  \includegraphics[width=0.45\textwidth]{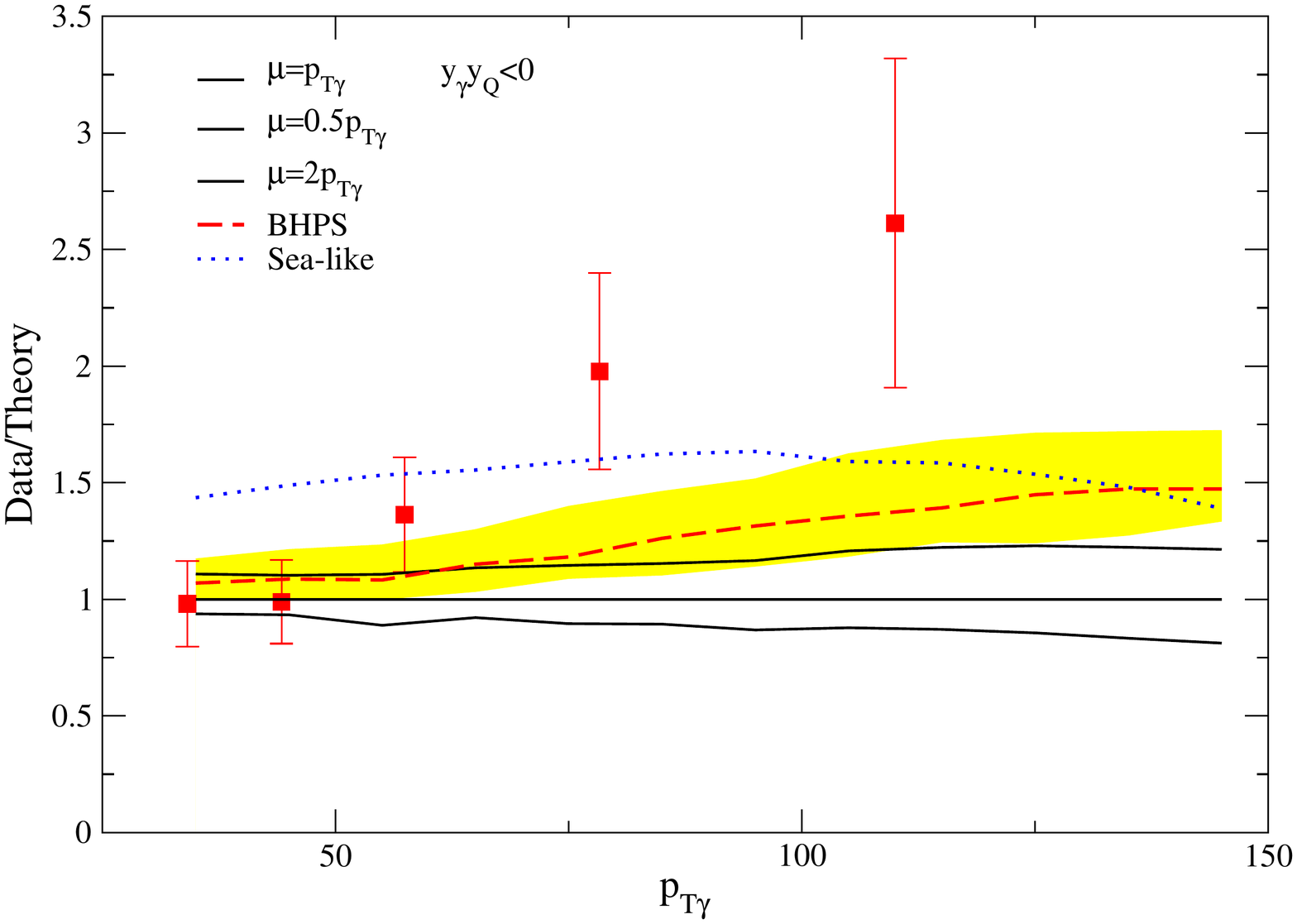}
  \caption{Data/Theory plot for NLO direct photon production in association with a charm quark jet at the Tevatron.}\label{fig:tevatron}
\end{figure}
the radiation subprocesses which include a light quark initial states as well. Especially at the Tevatron, these subprocesses can
even dominate the total cross-section (for details see \cite{Stavreva:2009vi}).
\section{Predictions for Tevatron, RHIC and the LHC}
Here we present predictions mostly in the form of differential cross-section over the transverse momentum of the photon for three 
different colliders - the Tevatron, RHIC and the LHC. We use specific cuts for minimal transverse momenta of the photon and the heavy 
quark jet, rapidity of the photon and we use isolation criteria for a direct photon specified by each experiment. All cuts are given in 
Tab.~\ref{tab:cuts}.

Data are already available for the process from the D0 experiment at the Tevatron and as can be seen in Fig.~\ref{fig:tevatron}, data 
at high transverse momentum of the photon do not agree with the standard NLO prediction and are also outside the uncertainty band 
shown in the Fig.~\ref{fig:tevatron}.
\begin{figure}[htb]
  \centering
  \includegraphics[width=0.45\textwidth]{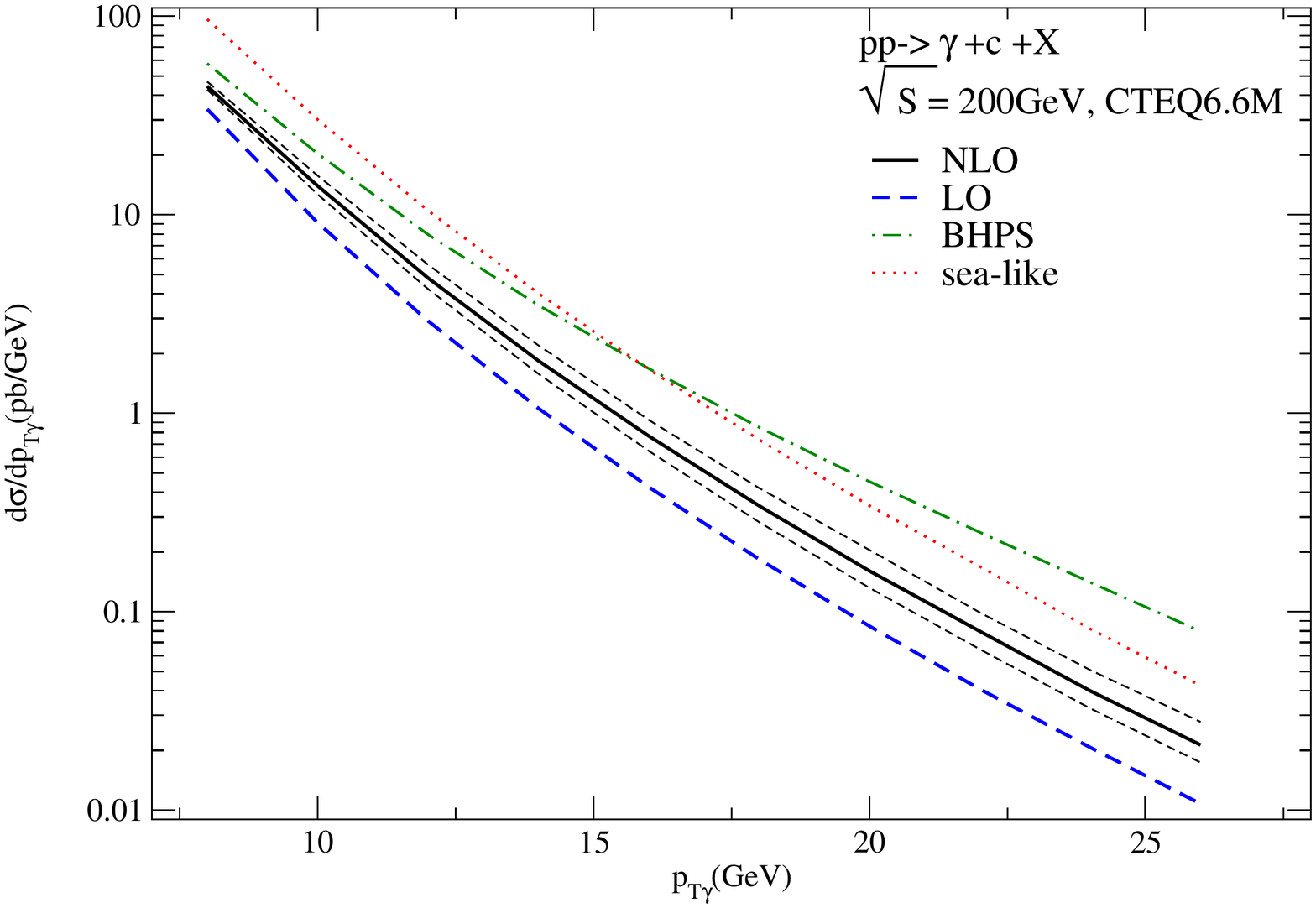}
  \includegraphics[width=0.34\textwidth,angle=90]{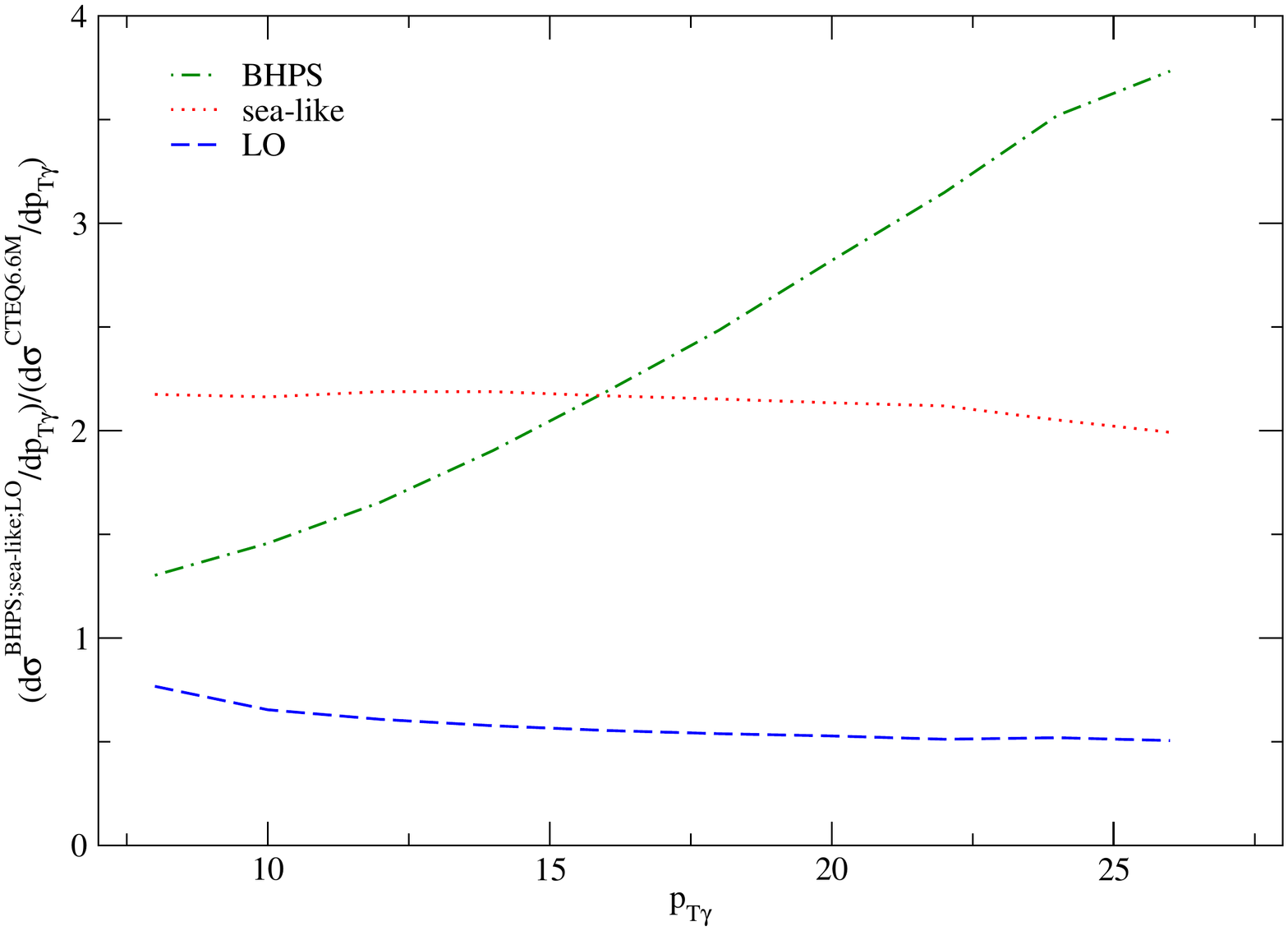}
  \caption{Differential cross-section for direct photon production with a charm quark jet at RHIC showing the LO prediction (blue dashed), 
NLO prediction with the uncertainty (solid black) and prediction for two models of intrinsic charm - BPHS (green dash-dotted), sea-like (red dotted).}
  \label{fig:rhic}
\end{figure}

\begin{table}[htb]
	\centering
\begin{tabular}{|lcccc|}
\hline 
Exp. &  & $p_T$ min.  & $y$  & Isolation\\
\hline 
\multirow{2}{*}{RHIC-Phenix} & photon      & 7 GeV  & $|y_\gamma|<0.35$ & $R=0.5$, $p_T=0.7$ GeV\\
                             & heavy quark & 5 GeV  & $|y_Q|<0.8$ & --- \\[1mm] \hline
\multirow{2}{*}{LHC-CMS}	 & photon      & 20 GeV  & $1.56<|y_\gamma|<2.5$ & $R=0.4$, $p_T=4.2$ GeV\\
                             & heavy quark & 18 GeV  & $|y_Q|<2.0$ & --- \\[1mm]
\hline
\end{tabular}
\caption{Kinematic cuts for all collider scenarios. \label{tab:cuts} }
\end{table}
The BHPS model of intrinsic charm enhances the cross-section at high $x$ which corresponds 
exactly to the region of large $p_T$ where data deviate from the theory prediction. Unfortunately not even the BHPS model can fully 
explain the data but it may indicate the presence of intrinsic component
of the charm quark PDF.

\begin{figure}[bht]
  \centering
  \includegraphics[width=0.45\textwidth]{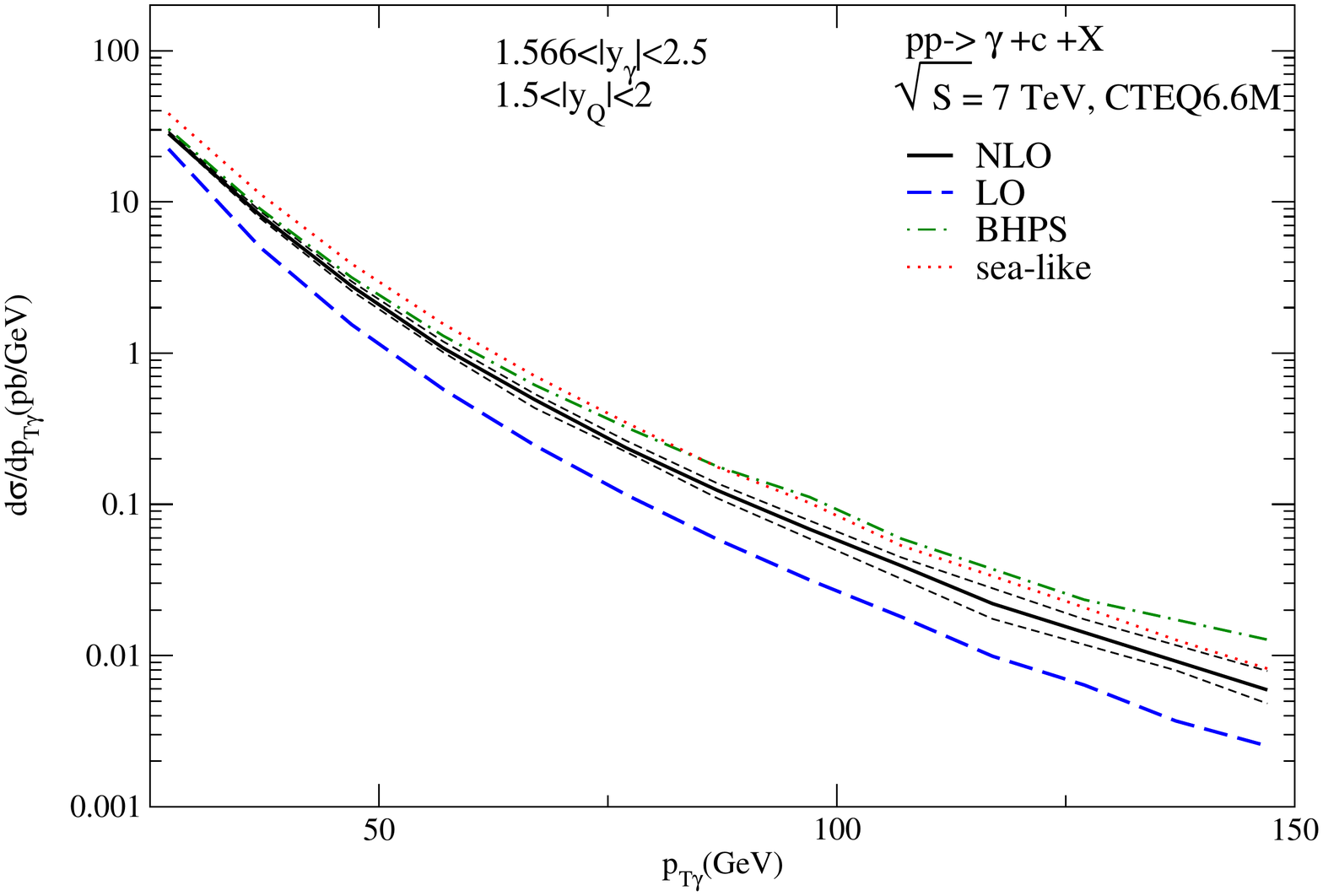}
  \includegraphics[width=0.45\textwidth]{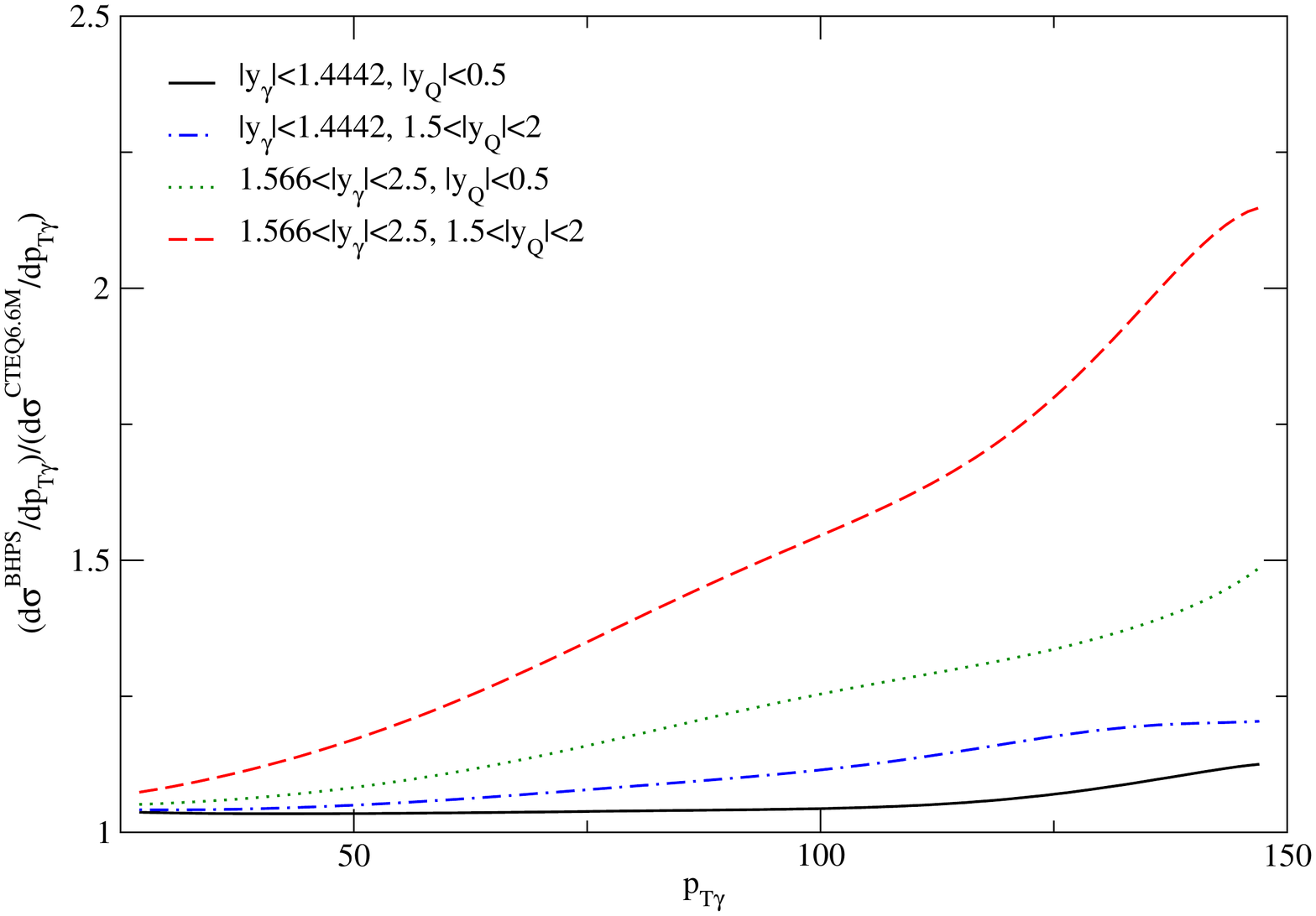}
  \caption{Differential cross-section for direct photon production with a charm quark jet at the LHC showing the same curves as in 
Fig.~\ref{fig:rhic}.}
  \label{fig:cms}
\end{figure}

A similar predictions to the one for Tevatron can be made for RHIC and the LHC. As can be seen from Figs.~\ref{fig:rhic}, \ref{fig:cms}, 
a possible measurement of the direct photon production with a heavy quark at RHIC is more sensitive to different models of
intrinsic charm because $\sqrt{s}$ at RHIC is smaller than at the LHC. That is why the direct photon production at RHIC probes a 
higher $x$ of the PDF of the colliding particles which is sensitive to the BHPS model of intrinsic charm.
\section{Conclusions}
We have showed predictions for direct photon production process in association with a heavy quark jet. Based on the discrepancy between data
and theory at NLO at the Tevatron, we focused on the case of a charm jet because the charm intrinsic contribution is less suppressed in 
comparison to a possible bottom quark intrinsic contribution. We have shown that using the studied process, one can test models of intrinsic 
charm such as the BPHS model which modify the charm PDF at large $x$ beyond the reach of DIS measurements. 


{\raggedright
\begin{footnotesize}


 \bibliographystyle{DISproc}
 \bibliography{kovarik_karol_hq.bib}
\end{footnotesize}
}


\end{document}